\documentclass[%
 reprint,
unsortedaddress,
 amsmath,amssymb,
 aps,
prb,
]{revtex4-1}

\usepackage[dvipdfmx]{graphicx}
\usepackage[dvipdfmx]{color}
\usepackage[breaklinks]{hyperref}
\hypersetup{colorlinks=true, linkcolor=blue, urlcolor=dblue, citecolor=blue}
\usepackage{dcolumn}

\usepackage{amsmath} \usepackage{amssymb}
\usepackage{accents} \usepackage{bbm}
\usepackage{mathrsfs}
\usepackage{accents}

\newcommand{\tr}[1]{\mathrm{Tr}\left[#1\right]}

\newcommand{\ut}[1]{\undertilde{#1}}

\newcommand{\ua}[0]{\uparrow} \newcommand{\da}[0]{\downarrow}

\newcommand{\gns}[0]{G_{\mathrm{NS}}} 
\newcommand{\gnn}[0]{G_{\mathrm{NN}}} 
\newcommand{\ngns}[0]{\bar{G}_{\mathrm{NS}}} 

\definecolor{dgreen}{RGB}{00, 120, 00}
\hypersetup{colorlinks=true, linkcolor=blue, urlcolor=blue, citecolor=blue, }
\usepackage{booktabs}

\usepackage{ulem}

\begin{document}
\preprint{APS/123-QED}

\title{Tunneling conductance of $d+ip$\,-wave superconductor}

\author{Yuhi Takabatake}
\author{Shu-Ichiro Suzuki}\email[Corresponding author: ]{suzuki@rover.nuap.nagoya-u.ac.jp}
\author{Yukio Tanaka}

\affiliation{Department of Applied Physics, 
Nagoya University, Nagoya 464-8603, Japan}

\date{\today}

\begin{abstract}
We theoretically investigate the tunneling conductance of the $d+ip$-wave 
superconductor that has recently been proposed to be realized at the (110) surface of
a high-$T_c$ cuprate superconductor. Utilizing the quasiclassical
Eilenberger theory, we obtain the self-consistent pair potentials
and the differential conductance of the normal-metal/$d+ip$-wave
superconductor junction. We demonstrate that the zero-bias peak 
of a $d$-wave superconductor is robust against the
spin-triplet $p$-wave surface subdominant order, even though it is fragile
against the spin-singlet $s$-wave one. Comparing our numerical results
with experimental results, we conclude that the spin-triplet $p$-wave surface
subdominant order is feasible. 
\end{abstract}

\pacs{}
\maketitle


\section{\label{sec:introduction} Introduction}

\begin{figure}[b!]
	\centering
        \includegraphics[width=0.46\textwidth]{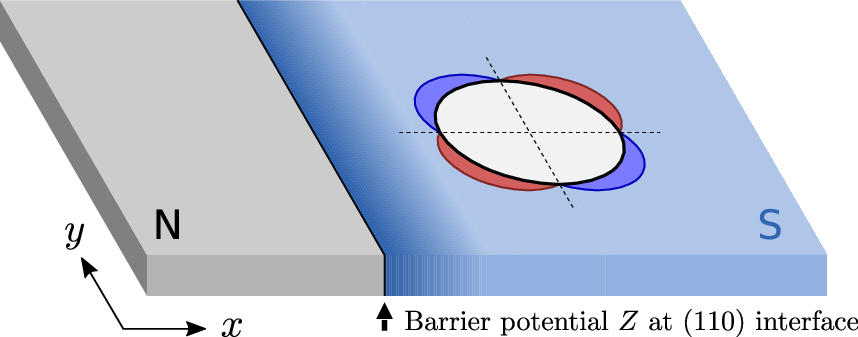}
	\caption{Schematic of the normal-metal (N)/ superconductor (S) junction.
	The $d_{xy}$-wave superconductor
	that corresponds to the (110) of a cuprate superconductor is realized in S.
	Because of the flat-band instability, a subdominant pair potential
	is induced near the interface. 
	There is a barrier potential $Z$ at the interface. }
	\label{fig:Sche}
\end{figure}

Unconventional superconductors (SCs) can host surface bound states, the
Andreev bound states (ABSs), forming a zero-energy flat band
\cite{ABS, ABSb, Hu, MS95, Kashiwaya00, ABSR2}. The zero-energy ABSs
can be observed as a zero-bias conductance peak (ZBCP) in the
quasiparticle tunneling spectra of the junctions of a normal metal and
a high-$T_{c}$ cuprate (i.e., a spin-singlet $d$-wave SC)
\cite{TK95,Experiment1, Experiment2, Experiment3, Experiment4,
Experiment5, Experiment6, Experiment7, Experiment8, Saadaoui13}. In
addition, the ABSs induce a Josephson current with a low-temperature
anomaly \cite{TK96b,TK97,BBR96} and a paramagnetic Meissner current
\cite{Higashitani_JPSJ_97, BKK00, Meissner3, Asano_PRL_2011,
Yokoyama_PRL_2011, Suzuki14, Higashitani_JPSJ_2014, Suzuki15,
Espedal_PRL_16, ShuPingLee_PRB_2017}. The origin of the ABSs has been
clarified from the view point of the topological invariant defined
using the bulk Hamiltonian \cite{STYY11}.


The zero-energy ABSs may be fragile against perturbations because of the
high degeneracy of the flat band. The surface $s$-wave subdominant order
originating this instability was proposed \cite{matsumoto951, kuboki96} in 1995. 
Theoretically, the subdominant
$s$-wave component splits the zero-energy
peak in the local density of states (LDOS) 
\cite{kashiwaya95calculation, Tanuma99, tanuma01d+is} and gives rise to a spontaneous 
surface current by breaking the time reversal symmetry
(TRS) \cite{matsumoto952}. 
In experiments, however, neither such peak splitting
nor TRS breaking has been observed \cite{Experiment1,
Experiment2, Experiment4, Experiment5, Experiment6, Experiment7,
Experiment8, Saadaoui13}, except for a few cases \cite{Experiment3, Fogel, Krupke99}.
More seriously, the induced $s$-wave pair potential requires an on-site attractive 
interaction, in contradiction to the strong repulsive 
interaction in the cuprate.
The instability of the ABSs is not still conclusive, even though other
possibilities have been pointed out, such as surface ferromagnetism
\cite{Potter14}, spin density wave \cite{honerkamp}, staggered flux
phase \cite{kuboki14, kuboki2015}, and translational symmetry breaking
\cite{Vorontsov09, higashitani15, Miyawaki15, miyawaki17,haakansson15, 
Holmvall18, Miyawaki18, Holmvall19}. 

The spin-triplet $p$-wave subdominant order has recently been proposed
using the finite-size two-dimensional Hubbard model with the random-phase
approximation \cite{matsubara20}. 
The ferromagnetic fluctuation caused by the ABSs can
stabilize such a $p$-wave subdominant order, which breaks the 
TRS \cite{matsubara20_2}. 
The obtained $d+ip$-wave pairing shows different properties compared 
with those of the $d+is$-wave: there is no clear zero-energy splitting in the LDOS 
and no spontaneous 
surface current. Although a number of papers have studied the $d+is$-wave
state \cite{matsumoto951,matsumoto952,Tanuma99,tanuma01d+is},  
the unique properties of the $d+ip$-wave state have not yet been 
clarified. In particular, the mixture of the spin-singlet and 
spin-triplet pairs would cause non-trivial phenomena. 

We here study the conductance spectra of the
normal-metal/$d_{xy}$-wave SC junctions with a subdominant $p_y$-wave 
pair potential at the interface and compare the results with those for the well-known 
$d+is$-wave superconducting junction. 
We consider a ballistic planar junction, as shown in Fig.~\ref{fig:Sche}, where the barrier potential 
is present at the interface.
Utilizing the quasiclassical Eilenberger formalism, we obtain
the differential conductance using the pair potential obtained by solving the
self-consistency equation. The calculated results show that the ZBCP 
of the $d_{xy}$-wave SC 
can survive under the spin-triplet $p_y$-wave subdominant
pair potential, even though it is fragile against the spin-singlet $s$-wave
subdominant pair potential. 
From the spin-resolved conductance spectra, we show that the transport properties of
the $d+ip$-wave junction strongly depend on the spin of an injected electron 
 because of the coexistence of the spin-triplet and singlet pairs near the interface. 

We also investigate whether the ZBCP can survive, even if the $p$-wave
surface attractive potential is short-range and strong, as pointed out in 
Ref.~\onlinecite{matsubara20}. The ZBCP is
demonstrated to be robust even against such a short-range attractive
potential. Comparing our results with experimental data, 
we conclude that our results, which take the subdominant $p$-wave order into account phenomenologically, support recent theoretical predictions based on microscopic calculations.


\section{Model and formulation}
\label{sec:Eilen}

We consider the ballistic normal-metal/$d_{xy}$-wave SC junction shown in 
Fig.~\ref{fig:Sche}, where the SC and the normal-metal (N) occupy $x \geq 0$ and $x<0$ respectively. 
In a ballistic SC, Green's function obeys the Eilenberger equation \cite{Eilenberger}: 
\begin{align}
  & i v_{F_x} \partial_{x} \check{g}_{\alpha\alpha} 
  = -\alpha [i\omega_n \check{\tau}_3+\check{\Delta}_\alpha, 
	\check{g}_{\alpha\alpha}], \label{eq:Eilenberger} \\
	& \check{g}_{\alpha\alpha}(\phi,x,i\omega_n) 
	= \left[ \begin{array}{rr}
	   \hat{    g }_{\alpha \alpha} &  \hat{    f }_{\alpha \alpha} \\
	  -\hat{\ut{f}}_{\alpha \alpha} & -\hat{\ut{g}}_{\alpha \alpha}
	\end{array} \right]
\end{align}
where $\hat{g}$ and $\hat{f}$ are the normal and anomalous Green's
functions, $v_{F_x}$ is the $x$ component of the Fermi velocity $v_F$, 
$\omega_n$ is the Matsubara frequency with the integer $n$, 
the direction of the momentum is characterized by the angle $\phi$ ($k_x = \alpha \cos \phi$, 
and $k_y = \sin \phi$ with $\alpha = \pm 1$ and $-\pi/2 \leq \phi 
\leq \pi/2$),  
$\check{\Delta}_\alpha = \check{\Delta}_\alpha(\phi,x)$, 
and 
$\check{\tau}_j$ ($j=1$, $2$, or $3$) are
the Pauli matrices in the particle-hole space. In this study, the symbols $\check{\cdot}$ and 
$\hat{\cdot}$ denotes the matrices in the particle–hole and spin space, respectively. 
The 
pair-potential matrix is defined as 
\begin{align}
  & \check{\Delta}_\alpha=
  \left[\begin{array}{cc}
      & \hat{\Delta}        \\
       -\hat{\Delta}^\dagger&
  \end{array}\right], 
	\hspace{6mm}
	%
	\hat{\Delta}=
  \left[\begin{array}{cc}
      & \Delta_{\ua \da} \\
        \Delta_{\da \ua} &
  \end{array}\right], 
\end{align}
where we have omitted the index $\alpha$. 
The spin structure of the pair potential is parametrized as 
\begin{align}
  \hat{\Delta} = 
	\left\{ \begin{array}{cc}
	  \Delta_d (i \hat{\sigma}_2)+ i\Delta_p \hat{\sigma}_1 & 
		\text{for $d+ip$ wave, }\\[2mm]
	  (\Delta_d + i\Delta_s)i \hat{\sigma}_2 & 
		\text{for $d+is$ wave, }
	\end{array} \right. 
  \label{eq:pair-spin}
\end{align}
where $\Delta_\mu$ ($\mu = s$, $p$, and $d$) is the amplitude of the
$\mu$-wave pair potential and $\hat{\sigma}_j$ ($j=1$, $2$, or $3$) are
the Pauli matrices in spin space. 
The momentum dependence of the pair potentials are given by 
\begin{align}
  \Delta_d (x, \phi) = & \Delta_d (x) \sin (2\phi), \\
  \Delta_p (x, \phi) = & \Delta_p (x) \sin  \phi, \\
  \Delta_s (x, \phi) = & \Delta_s (x). 
\end{align} 

The pair potentials are determined by the self-consistency 
equation: 
\begin{align}
  \label{pair potential'}
  &\Delta_\mu = \Lambda_\mu 
	\sum_{n=0}^{n_c}
	\left \langle
  \mathrm{Tr}[\hat{V}_\mu(\phi) 
	\hat{f}_{\alpha \alpha}(\phi,x,i\omega_n)]
	\right \rangle_{\mathrm{FS}}, 
\end{align}
where the angular brackets means the angle average on the Fermi
surface: 
  $ \langle \cdots \rangle = \sum_{\alpha}
  \int_{-\pi/2}^{\pi/2}  \cdots ({d \phi}/{2\pi})$ and $n_c$ is the
cutoff integer which is decided by the relation $2n_c+1 < \omega_c /
\pi T \leq 2n_c+3$ with $\omega_c$ being the cutoff energy. 
The attractive potential depends on the pairing symmetry: 
\begin{align}
  \hat{V}_\mu(\phi) = 
	\left\{ \begin{array}{ll}
	  2 i \hat{\sigma}_2 \sin (2 \phi) & \text{for the $d$ wave, }\\[2mm]
	  2   \hat{\sigma}_1 \sin    \phi  & \text{for $p$ wave, }\\[2mm]
	    i \hat{\sigma}_2                 & \text{for $s$ wave. }
	\end{array} \right. 
\label{}
\end{align}
The coupling constant $\Lambda_\mu$ is 
\begin{align}
 \Lambda_\mu = 2\pi T
 \left [ \ln
 \left(\frac{T}{T_\mu}\right)+\sum_{n=0}^{n_c}\frac{1}{n+{1}/{2}}
 \right]^{-1}, 
\end{align}
where $T_\mu$ is the effective critical temperature. Namely, the ratio
$T_p/T_d$ and $T_s/T_d$ characterize the
amplitude of the subdominant pair potential. 

The microscopic theories\cite{matsubara20, matsubara20_2} suggest that the attractive
potential for the $p$-wave channel may be short-range. To model such a
short-range potential, we introduce the spatial-dependent attractive
potential $\Lambda_\mu'(x)$ as 
\begin{align}
  \Lambda_\mu'(x) 
	= 
  \Lambda_\mu \exp \left[ -x/\kappa \right], 
  \label{eq:kappa}
\end{align}
where $\kappa$ is the decay parameter which characterizes the length
scale of the attractive potential. Replacing $\Lambda_\mu$ in
Eq.~\eqref{pair potential'} by $\Lambda_\mu '(x)$, we can
self-consistently calculate the subdominant pair potential under the
short-range attractive potential.

When the pair-potential matrix does not have the diagonal component as in 
Eq.~\eqref{eq:pair-spin}, the $4 \times 4$ Eilenberger equation can be
decomposed into two $2 \times 2$ ones: 
\begin{align}
  & i v_{F_x} \partial_{x} \tilde{g}^X_{\alpha\alpha} 
  = -\alpha [i\omega_n \tilde{\tau}_3+\tilde{\Delta}^X_\alpha, 
	\tilde{g}^X_{\alpha\alpha}]. 
\end{align}
The spin-reduced Green's function $\tilde{g}^X$ and the pair potential are 
defined as 
\begin{align}
  & \tilde{g}^O=
  \left[\begin{array}{cc}
           g_\ua  &      f_{\ua \da} \\
      -\ut{f}_{\da \ua} & -\ut{g}_\da 
  \end{array}\right], 
	\hspace{5mm}
    \tilde{g}^I=
  \left[\begin{array}{cc}
           g_\da  &      f_{\da \ua} \\
      -\ut{f}_{\ua \da} & -\ut{g}_\ua 
  \end{array}\right], 
	\\
  & \tilde{\Delta}_\alpha^ O=
  \left[\begin{array}{cc}
      & \Delta  _{\ua \da} \\
       -\Delta^*_{\ua \da} &
  \end{array}\right], 
	\hspace{3mm}
    \tilde{\Delta}_\alpha^ I=
  \left[\begin{array}{cc}
      & \Delta  _{\da \ua} \\
       -\Delta^*_{\da \ua} &
  \end{array}\right]. 
\end{align}
where we omit the direction index $\alpha$ and the index $X=O$ and $I$ 
means the outer and inner components in the Nambu space respectively. 
Hereafter we make the index $X$ explicit only when necessary. 
Using the Riccati parametrization  \cite{Schopohl_PRB_95,
Eschrig_PRB_00, Eschrig_PRB_09}, the quasiclassical Green's
functions $\hat{g}^X_{\alpha \alpha}$ can be expressed as 
\begin{align}
\label{gpp}
	\tilde{g}_{\alpha \alpha}=
	\frac{i}{1-D_\alpha F_\alpha}
	\left(\begin{array}{cc}
	1+D_\alpha F_\alpha&     2i\alpha F_\alpha\\[1mm]
	         2i\alpha D_\alpha&-(1+D_\alpha F_\alpha)
	\end{array}\right)
\end{align}
where $D_\alpha = D_\alpha^X$ and $F_\alpha = F_\alpha^X$ are the
so-called Riccati amplitudes. 
The Riccati amplitudes obey the following Riccati-type differential equations: 
\begin{align}
	& \label{Dp} v_{F_x}\partial_{x} D_+ = 2\omega_n D_+ +\Delta  _+D^2_+ -\Delta^*_+,\\
	& \label{Dm} v_{F_x}\partial_{x} D_- = 2\omega_n D_- +\Delta^*_-D^2_- -\Delta  _-,\\
	& \label{Fp} v_{F_x}\partial_{x} F_+ =-2\omega_n F_+ +\Delta^*_+F^2_+ -\Delta  _+,\\
	& \label{Fm} v_{F_x}\partial_{x} F_- =-2\omega_n F_- +\Delta  _-F^2_- -\Delta^*_-, 
\end{align}
The Eilenberger equation is supplemented by the boundary conditions
\cite{AAHN_PRB_89, Nagato_JLTP_93, Eschrig_PRB_00, Tanuma_PRB_01, Hirai_PRB_03, Eschrig_PRB_09}. The boundary conditions at the N/SC interface are given by: 
\begin{align}
	F_\pm(0)=-R^{s_\omega}D_\mp(0),
  \label{kyoukai1}
\end{align}
where $R$ is the reflection probability amplitude and $s_\omega = \mathrm{sgn}[\omega_n]$.

\subsection{Real-energy representation}
In order to discuss the quantities depending on the energy, we need
the Green's function in the real-frequency representation. The Green's
function in the real-energy space can be obtained by the analytic
continuation; $i\omega_n \to E + i \delta$. In this case, the Riccati
amplitudes are also converted as 
\begin{align}
	D_{\pm}^X(x, i \omega_n )=i\Gamma^X_{\pm}(x, E), \quad
	F_{\pm}^X(x, i \omega_n )=i \zeta^X_{\pm}(x, E). 
\end{align}
From the Riccati amplitude $\Gamma_\pm$, we can directly calculate the
angle-resolved conductance as a function of $E=eV$ and $\phi$: 
\begin{align}
\label{sigmar}
	\sigma_{R}
	=
	\frac{1+\sigma_N |\Gamma_{+}|^2+(\sigma_{N}-1)|\Gamma_{+}\Gamma_{-}|^2}
	{|1+(\sigma_{N}-1)\Gamma_{+}\Gamma_{-}|^2}
\end{align}
where $\Gamma_{\pm}=\Gamma_{\pm}(x=0, E,\phi)$, $\sigma_{R}=\sigma_{R}(E,\phi)$, $\sigma_{N}=\sigma_{N}(\phi)$, 
$V$ is the bias voltage applied to the junction and $\Gamma_\pm = \Gamma_\pm^X$. 
The conductance in the normal state is obtained by solving the
scattering problem: $ \sigma_{N}(\phi)=1-R={\cos^2\phi}/(Z^2 +\cos^2 \phi) $, 
where we assume the potential barrier $Z v_F \delta(x)$ with $\delta(x)$ being 
the delta function. 

The total conductance is defined as 
\begin{align}
	& \gns(E)=
	\sum_{X}
	\int^{\pi/2}_{-\pi/2}
	\gns'^X (E, \phi)
	\cos\phi d\phi, 
	\label{eq:gns}
	\\
	& \gns'^X (E, \phi) =
	\sigma_{N}(\phi)\sigma^X_{R}(E,\phi). 
\end{align}
where $\gns'^X(E, \phi)$ is the angle-resolved differential
conductance. The conductances $\gns'^O(E, \phi)$ and  $\gns'^I(E, \phi)$ correspond to those to 
up-spin and down-spin injections, respectively. It is convenient to introduce the normalized 
conductance
$\ngns = \gns/\gnn$ with $
	\gnn =
	2 \int^{\pi/2}_{-\pi/2}
	\sigma_N
	\cos\phi d\phi
$. 

The local density of states (LDOS) can be obtained from the
quasiclassical Green's function. The LDOS is given by 
\begin{align}
  & \rho = \frac{1}{\pi}\int_{-\pi/2}^{\pi/2} \rho'(\phi) {d\phi},
	\quad
	  \rho'(\phi) =
	\frac{1}{2} \sum_{\alpha}
  \mathrm{Tr}[
	\check{g}_{\alpha \alpha}]. 
\end{align}
In quasiclassical theory, the LDOS is normalized by its normal-state value.

\section{Results}
\label{sec:DC}

\subsection{Differential conductance}
\begin{figure}[tb]
	\centering
  \includegraphics[width=0.48\textwidth]{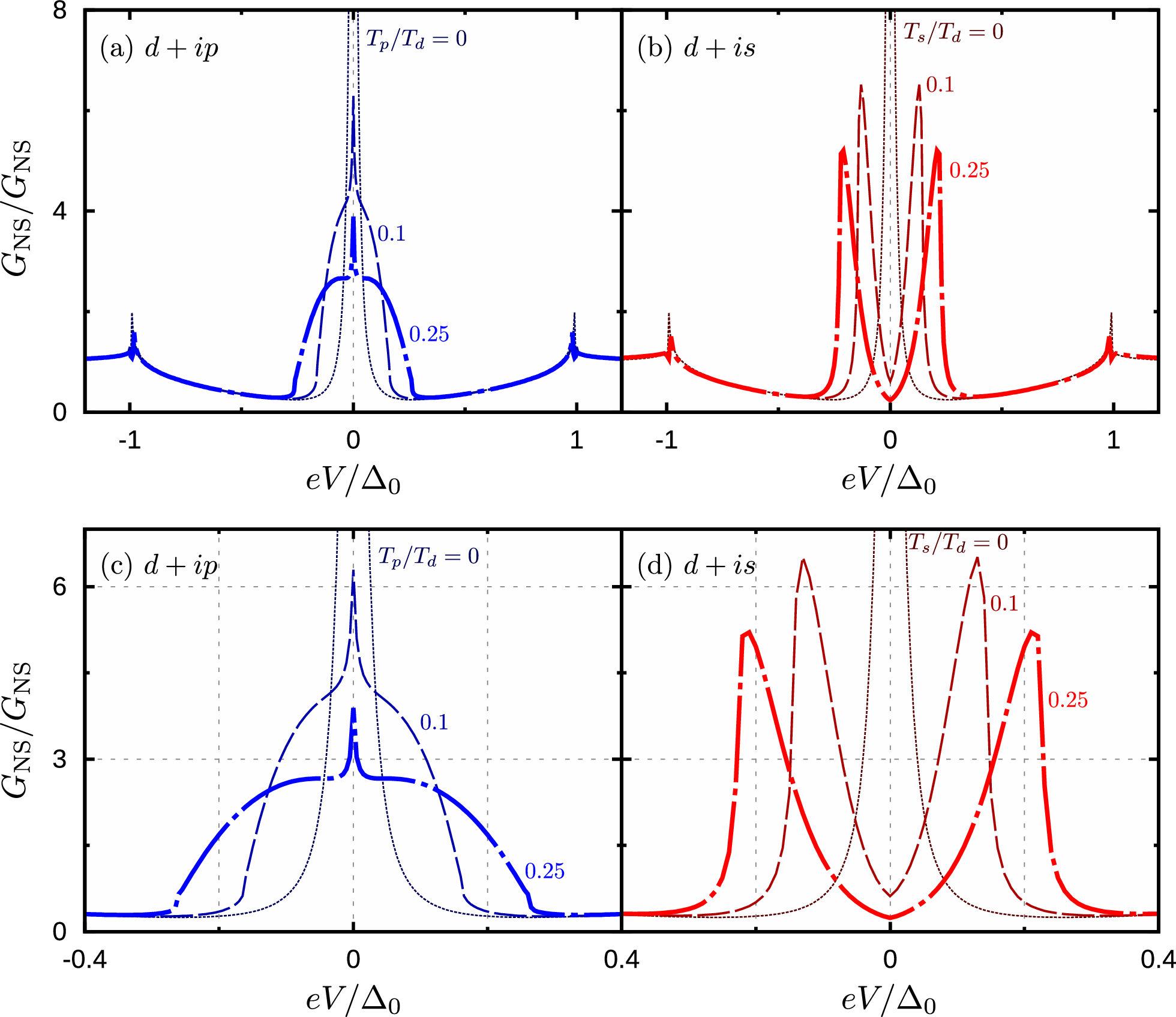}
	\caption{Differential conductances of (a) $d+ip$- and (b) $d+is$-wave
	junctions. The magnified figures are shown in (c) and (d). 
	The ZBCP is not split by the $p$-wave subdominant
	component but by the $s$-wave one. The ZBCP is robust against the
	$p$-wave subdominant pair potential but is fragile against the
	$s$-wave one. 
	The barrier potential is set to $Z=3$. 
	The pair potential is determined self-consistently. The temperature
	and the cutoff energy 
	are  set to $T = 0.05T_d$ and $\omega_c = 2 \pi T_d$. 
	}
	\label{fig:Conduc}
\end{figure}

\begin{figure}[t]
	\centering
  \includegraphics[width=0.48\textwidth]{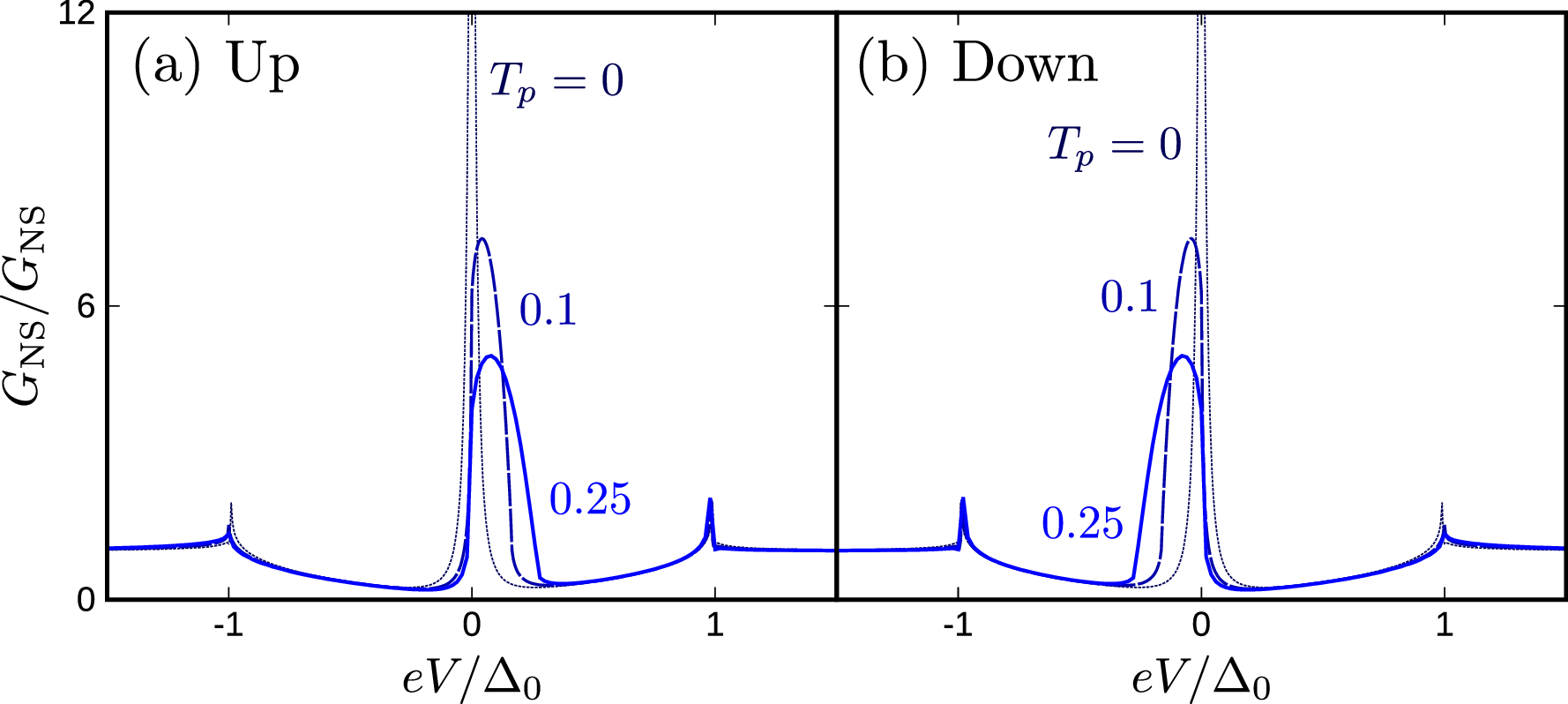}
	\caption{Injected-spin dependence of $\gns$ of 
	$d+ip$\,-wave junction The injected particle is assumed to be (a)
	up and (b) down, respectively. The differential conductance of the
	$d+ip$\,-wave junction depends
	on the injected spin because the subdominant component is
	spin-triplet pairing. The parameters are set to the same values used
	in Fig.~\ref{fig:Conduc}. 
	}
	\label{fig:Conduc_spin}
\end{figure}

\begin{figure*}[t]
	\centering
  \includegraphics[width=0.98\textwidth]{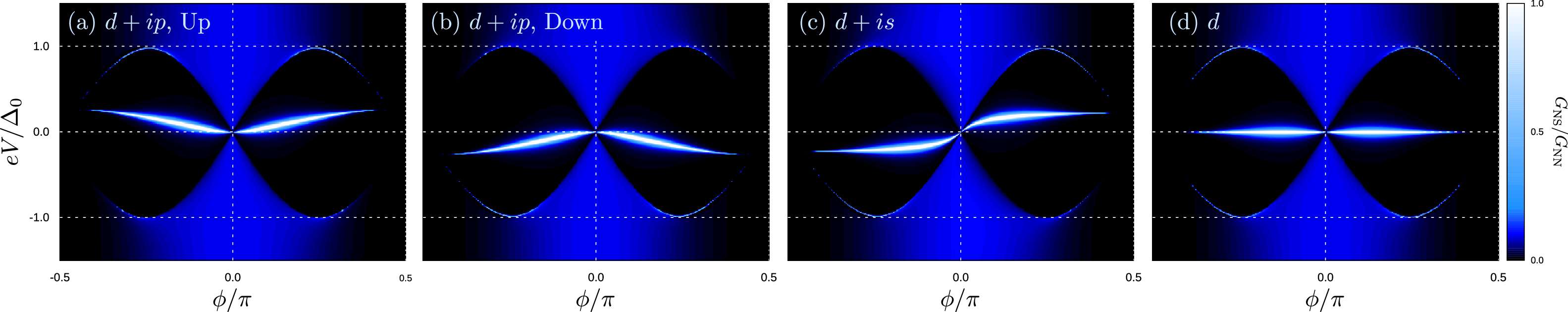}
	\caption{Angle-resolved differential conductances of $d+ip$\,-wave
	junctions for (a) up- and (b) down-spin injection. Zero-energy flat band of a $d$-wave
	SC becomes dispersive by a $p$\,-wave subdominant pair potential. 
	The dispersion of the bound states changed from a flat band to the
	V-shaped. 
	The surface states around $\phi=0$ stay around zero energy because 
	the $p$\,-wave component is small around $\phi=0$. The effective 
	critical temperatures for the subdominant components are set to $T_p = T_s = 0.25 T_d$ in (a), (b), and (c). }
	\label{fig:ARConduc-p}
\end{figure*}

The differential conductance for the $d+ip$- and $d+is$-wave junctions
are shown in Figs.~\ref{fig:Conduc}(a) and \ref{fig:Conduc}(b),
respectively, where the magnified figures are shown in 
Figs.~\ref{fig:Conduc}(c) and \ref{fig:Conduc}(d) and 
the pair potentials are determined
self-consistently (see Appendix for details). Throughout
this paper, the temperature and cutoff-energy are set to $T=0.05T_d$
and $\omega_c = 2 \pi T_d$. 
Figure~\ref{fig:Conduc} shows that the ZBCP is robust against the
$p$-\,wave subdominant component but fragile against the $s$-wave component. 
In the $d+ip$ case, the ZBCP survives even when $T_p/T_d = 0.25$. With
increasing $T_p/T_d$, the ZBCP becomes broader where the peak width is
roughly characterized by $T_p$. When $T_p/T_d = 0.25$, 
Two peaks seem to overlap at $eV=0$: sharper one and broader one. 
In the $d+is$ case, the ZBCP is fragile against the subdominant pair
potential as shown in Fig.~\ref{fig:Conduc}(b). In the presence of the
$s$-wave pair potential, the ZBCP is split into two peaks, where
distance between two peaks are characterised by $T_s$. This result is
consistent with Ref.\onlinecite{matsumoto951,matsumoto952}. 

We explain the origin of the sharp and narrow peaks for the
$d+ip$\,-wave junction by analyzing the injected-spin dependence of
$\gns$. The conductance for the up-spin and down-spin injections are
shown in Fig.~\ref{fig:Conduc_spin}(a) and \ref{fig:Conduc_spin}(b),
respectively, where the parameters are set to the same values used in
Fig.~\ref{fig:Conduc}. The center of the zero-energy peak for the
up-spin (down-spin) injection shifts from $E=0$ to a finite positive
(negative) energy, where the peak becomes broader simultaneously.
Although the peak center moves from the zero bias, the zero-energy
conductance $(\gns/\gnn)|_{E=0}$ always has an amplitude larger than
the
unity independent of the injected spin. Therefore, the ZBCP of the total
conductance [Fig.~\ref{fig:Conduc}(a)] does not disappear but becomes 
thicker by the subdominant $p$-wave order parameter. The spin-resolved
conductance for the $d+is$-wave junction (not shown) does not depend
on the injected spin because both of the $d$- and $s$-wave pairs are
spin singlet. 

The angle-resolved differential conductance are shown in
Fig.~\ref{fig:ARConduc-p}, where the pairing symmetry is assumed
$d+ip$-wave in (a) and (b), $d+is$-wave in (c), and pure $d$-wave in
(d). 
The injected spin is assumed up in Figs.~\ref{fig:ARConduc-p}(a), 
\ref{fig:ARConduc-p}(c) 
\ref{fig:ARConduc-p}(d), whereas 
down in Fig.~\ref{fig:ARConduc-p}(b), where 
the conductance for the $d+is$- and pure $d$-wave junctions does not
depend on the injected spin.
In the absence of a subdominant pair potential, the angle-resolved conductance $\gns'
(\phi)$ has a peak at the zero bias voltage (i.e., $eV=0$) independent of $k_y = k_F \sin \phi$, 
[see Fig.~\ref{fig:ARConduc-p}(d)].

\begin{figure}[tb]
        \includegraphics[width=0.48\textwidth]{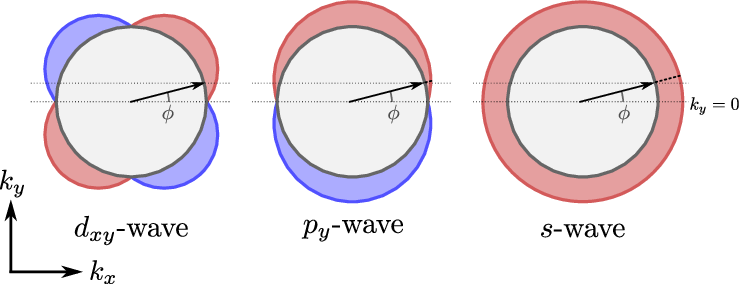}
	\caption{Schematic of the pair potentials. The $p_y$-wave subdominant component 
	has a small amplitude for the low-angle injection ($\phi \sim 0 $), where 
	the amplitude of the $s$-wave component is independent of the angle. Reflecting the $p_y$wave 
	nature, the zero-energy conductance peak of the $d+ip$-wave superconductor does not split. }
	\label{fig:Sche_pp}
\end{figure}

In the presence of the $p$-wave subdominant component, the in-gap
conductance peak changes from a flat band to the V
[inverted V] shape for up-spin [down-spin] injection, as shown in
Fig.~\ref{fig:ARConduc-p}(a) [Fig.~\ref{fig:ARConduc-p}(b)]. However, 
the in-gap peak around $\phi=0$ stay around $eV=0$ because 
the $p$\,-wave component has nodes at $\phi=0$ (see Fig.~\ref{fig:Sche_pp}). As a result, for
both spin injections, 
the zero-energy conductance have relatively large amplitude and the
ZBCP in the total conductance can survive even with the
$p_y$-wave subdominant component 

In the $d+is$-wave case, the in-gap conductance changes from the flat
band to a s-shaped one as shown
in Fig.~\ref{fig:ARConduc-p}(c), where the conductance does not
depend on the injected spin. Even around $\phi=0$, the ABSs are lifted from $eV=0$ because the $s$-wave pair
potential does not have any node on the Fermi surface  (see Fig.~\ref{fig:Sche_pp}). Reflecting this nodeless 
structure, the slope of the in-gap conductance peak around $\phi=0$ 
is much larger than those for the $d+ip$-wave junction. 
As a result, the conductance at $eV=0$ of the $d+is$-wave junction
is greatly reduced by the subdominant component. 

\begin{figure}[t]
	\centering
  \includegraphics[width=0.48\textwidth]{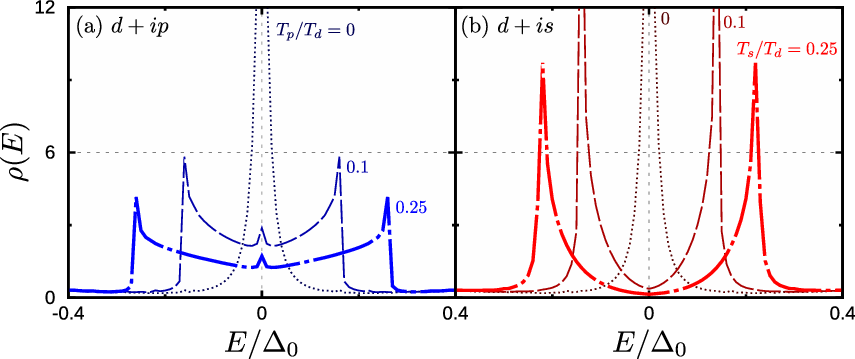}
	\caption{Local density of states at the interface of (a) $d+ip$\,-
	and (b) $d+is$-wave junctions. The results are normalized to its
	normal-state value. Differing from the differential
	conductance, the ZBCP in the LDOS is split by the subdominant pair
	potential regardless of the pairing symmetry of the subdominant pair
	potential. The parameters are set to the same values used in
	Figs.~\ref{fig:Conduc}(a) and \ref{fig:Conduc}(b), respectively. }
	\label{fig:SDOS}
\end{figure}

Contrary to the differential conductance, the LDOS reflects the
splitting of the zero-energy peak owing to the subdominant pair potential. The LDOS
at the interface of the $d+ip$\,- and $d+is$-wave junctions are shown
in Figs.~\ref{fig:SDOS}(a) and \ref{fig:SDOS}(b). Comparing
Figs.~\ref{fig:SDOS}(a) and \ref{fig:Conduc}(a), we see that the zero-energy peak 
splits even in the $d+ip$\,-wave junction, where the ZBCP does not split. 
The zero-energy peak moves to $E \sim \pm T_p$. 
In $\gns$, the surface state with
the small angle (i.e., $k_y \ll 1$) contributes more than those
with large angles (i.e., $|k_y| \sim  k_F$) because the conductance 
represents the electric current flowing in the $x$-direction; the more perpendicular
injection has the more contribution to the conductance (see the factor
$\cos \phi$ in the angle integration in Eq.~\ref{eq:gns}). On the
other hand, the LDOS is not directly related to the transport. Thus,
the channels with large angles also contribute to the LDOS. 
Although there are two high peaks in the LDOS of  the $d+ip$\,-wave
junction, a sharp but low peak appears at $E=0$. 
The origin of this low peak is the same as in the conductance. The
center of the LDOS peaks shift to the positive or negative side depending on
the spin. However, the LDOS for both spin
have a relatively large amplitude at $E=0$ and make a zero-energy peak in the
total LDOS. 

The spontaneous edge currents in the $d+ip$- and $d+is$-wave
SCs are explained in Appendix B with focusing on the symmetry of
the quasiclassical Green's function. The spontaneous current is absent
(present) in the $d+ip$-wave ($d+is$-wave) SC. 


\subsection{Barrier-strength dependence}

\begin{figure}[t]
  \includegraphics[width=0.48\textwidth]{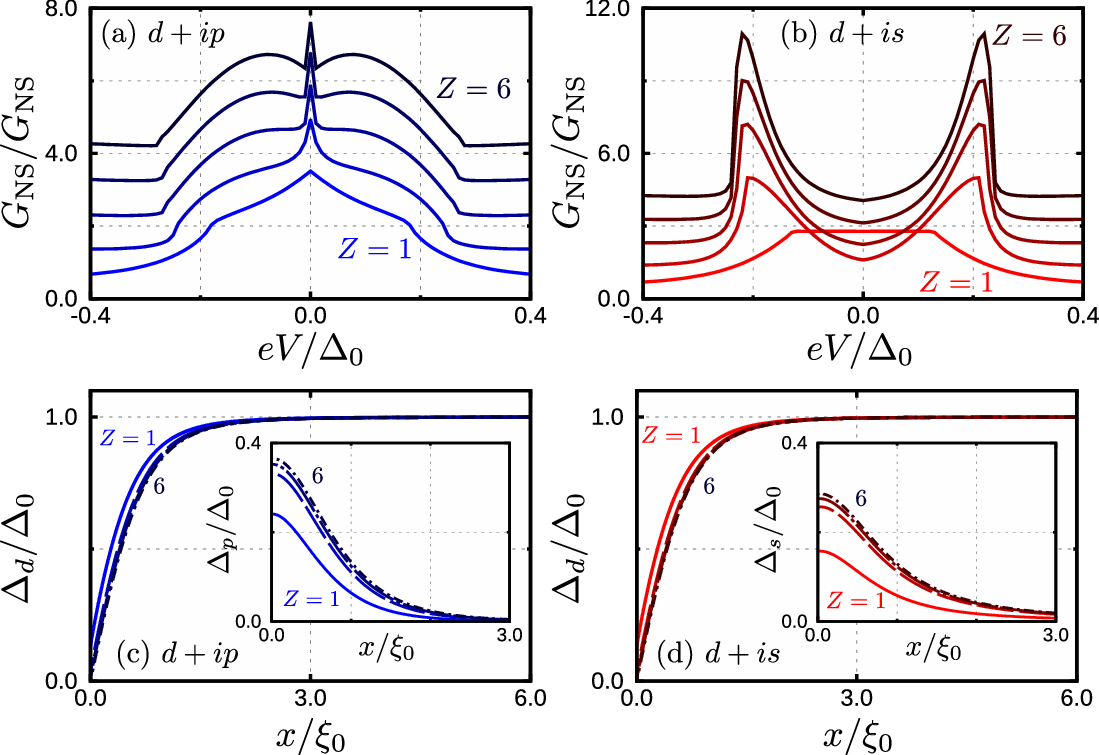}
	\caption{Evolution of the zero-bias conductance peaks 
	for (a) $d+ip$- and (b) $d+is$-wave junctions. In (a) and (b), the barrier potential is set to. 
	$Z=1$, 
	$2$, 
	$3$, 
	$4$, and 
	$6$. The calculated results are plotted with a shift of $\gnn$ with increasing $Z$. 
	The ZBCP for the $d+ip$-wave junction does not split regardless of the barrier strength, 
	whereas the ZBCP for the $d+is$-wave junction changes from a peak to finite-energy double peaks when $Z \geq 2$. 
	The spatial profiles of the pair potentials are shown in (c) and (d), 
	where the subdominant components are plotted in the insets. In (c) and (d), 
	the barrier potential is set to 	$Z=1$, 
	$2$, 
	$3$, and 
	$6$. 
	The effective critical temperature for the subdominant components are 
	set to 	$T_p/T_d = T_s/T_d = 0.25$. }
	\label{fig:Zdep}
\end{figure}
The differential conductance in the presence of a subdominant
component depends on the strength of the barrier potential
\cite{Tanuma2001}. The evolution of the ZBCP is shown in
Figs.~\ref{fig:Zdep}(a) and \ref{fig:Zdep}(b), where the $d+ip$- and
$d+is the $-wave superconductor assumed in Figs.~\ref{fig:Zdep}(a), and
\ref{fig:Zdep}(b) respectively. The barrier potential $Z$ changes asfollows: 
$Z=1$, $2$, $3$, $4$, and $6$. Effective critical temperature for
the subdominant components are set to $T_p/T_d = T_s/T_d = 0.25$. The
ZBCP for the $d+ip$-wave junction is present regardless of the
strength of the barrier parameter $Z$. We have confirmed that the ZBCP
does not split under even higher barrier potentials. In general, the
position of the midgap conductance peak is not influenced by the
barrier parameter $Z$. The larger barrier just results in the sharper
spectra. Therefore, the low-angle contribution as discussed above can
survive even with a large barrier potential. 

The zero-bias conductance for the $d+is$-wave is more sensitive to the
barrier potential than that for that for the $d+ip$-wave junction.
The amplitude of the zero-bias conductance reduces significantly with
increasing $Z$. Even with a rather small barrier potential (e.g.,
$Z=2$), the two peaks appear at $eV=0.25 \Delta_0$, which corresponds
to the amplitude of the subdominant $s$-wave pair potential. Namely,
the split peak would be observed more frequently in high-$T_c$
superconductor junctions if the $s$-wave subdominant order is
realized. 

The barrier-potential dependence of the pair potentials for the
$d+ip$- and $d+is$-wave junctions are shown in Figs.~\ref{fig:Zdep}(c)
and \ref{fig:Zdep}(d), where the $p$- and $s$-wave subdominant
components are shown in the insets. The barrier potential is set to
$Z=1$, $2$, $3$, and $6$. The dominant $d$-wave pair potential is
not strongly depend on $Z$. Their profiles for $Z \geq 2$ are almost
the same regardless the pairing symmetry of the subdominant
components. The amplitudes of both of the subdominant $p$- and
$s$-wave pair potentials increase with an increase in $Z$. The larger $Z$
generates the more subdominant components reflecting the parity mixing
by the inversion symmetry breaking.

\subsection{Effects of short decay length}

\begin{figure}[t]
  \includegraphics[width=0.48\textwidth]{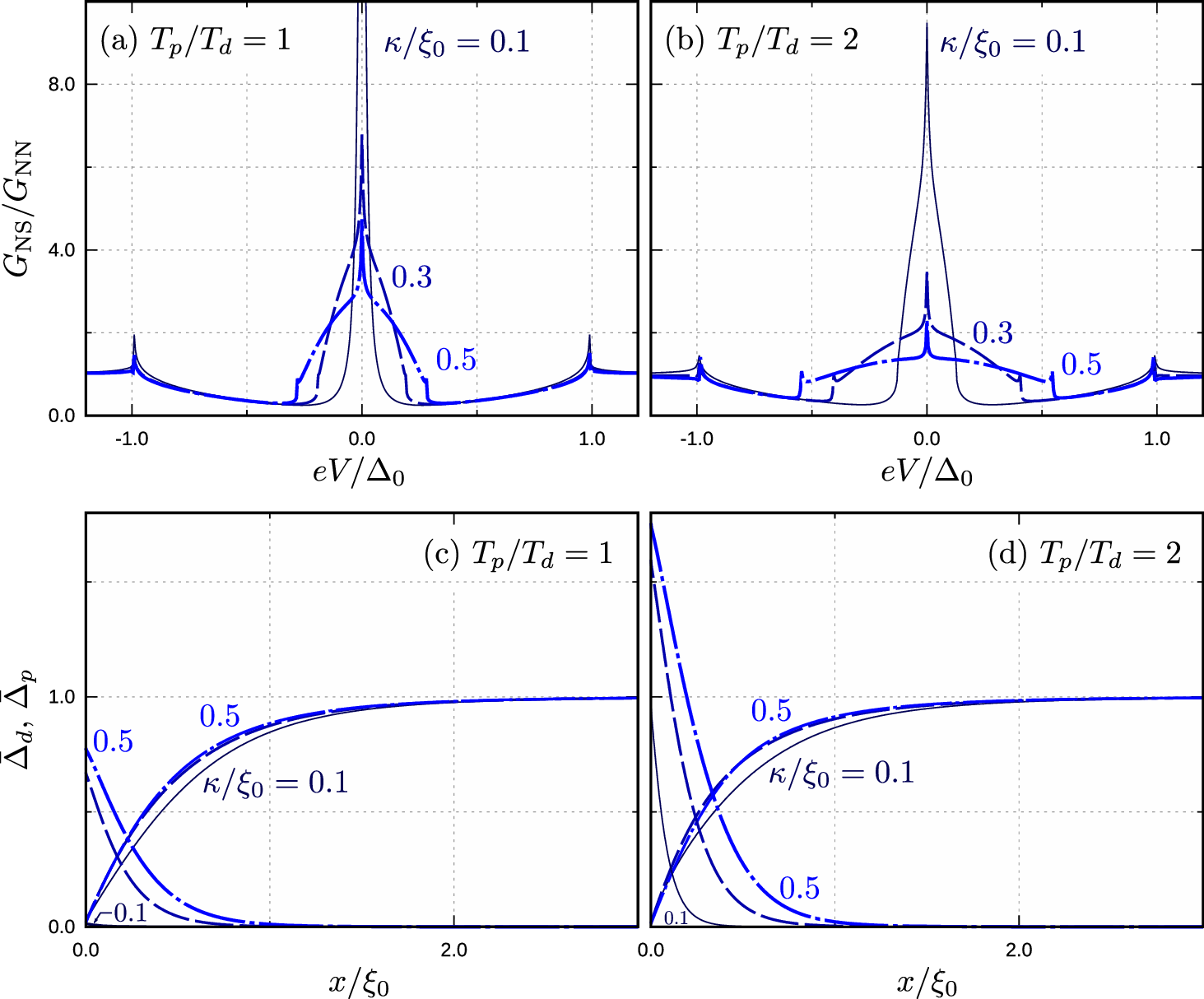}
	\caption{Effect of short decay length of the subdominant component.
	The differential conductances and the profile of the pair potentials
	are shown in top and bottom panels respectively, where 
	$T_p/T_d = 1$ in (a) and (c), and $T_p/T_d = 2$ in (b) and (d), respectively. 
	We see that the ZBCP is present even when $T_p > T_d$ and $\kappa <
	\xi_0$. }
	\label{fig:kappa}
\end{figure}

The microscopic calculations\cite{matsubara20, matsubara20_2} indicates 
that the $p$-wave attractive
interaction may be very strong in the very vicinity of the interface. The decay
length of surface $p$-wave component may be much shorter than the superconducting coherence length
and $T_p$ may be larger than $T_d$.
Such a short-range strong attractive interaction can be modeled by increasing
$T_p$, and by introducing the decay parameter $\kappa$ [see Eq.~\eqref{eq:kappa}]: 
The differential conductance and pair potentials with a short-range strong attractive
potential is shown in Figs.~\ref{fig:kappa}(a) and \ref{fig:kappa}(b).
The results 
show that the ZBCP is not split by the subdominant pair potential even when the attractive
potential is much larger than that for the dominant $d$-wave. With increasing decay
parameter, the ZBCP becomes broader because the influences from the
subdominant potential becomes larger.

\section{Discussion}
We have confirmed that our conclusions on the ZECP do not
depend on the details of the potential at the interface. We have
replaced the $\delta$-function insulating barrier with the rectangular
one. The results with the rectangular potential (not shown) are
qualitatively the same as those with the $\delta$-function one [e.g.,
Fig.~\ref{fig:Zdep}(a)]; the ZBCP of the $d+ip$-wave junction does not split. Therefore,
our conclusions are valid even when the interface potential has a
finite width (i.e., more realistic model than the $\delta$-function
barrier model). 

The conductance spectra of cuprate superconductors observed in
experiments to date have a peak at the zero energy \cite{Experiment1,
Experiment2, Experiment4, Experiment5, Experiment6, Experiment7,
Experiment8}, even though several
experiments have reported splitting of the ZBCP
\cite{Experiment3, Fogel, Krupke99}.
Our theoretical study, where the $p$-wave surface attractive
interaction is phenomenologically taken into account, demonstrates
that the $p$-wave subdominant pair potential does not split the
zero-bias peak. In particular, our results show that the microscopic theory
\cite{matsubara20, matsubara20_2} can be consistent with experimental results obtained
to date. 

The conductance spectrum of a $d+ip$-wave junction depends on the
spin of the injected particle because of the mixture of the $p$-wave spin-triplet and 
$d$-wave spin-singlet pairs.
Replacing the normal-metal electrode with a ferromagnetic metal will
provide useful information for detecting the surface subdominant
pair potential. Investigating how the $p$- and $s$-wave subdominant pair potentials 
modify the conductance spectra would be interesting. 

In this paper, we have studied the transport property of the $d+ip$-wave junction in
the ballistic limit by calculating the tunneling conductance. 
The induced $p$-wave pair, however, is more fragile
against impurity scatterings than are $s$-wave pairs \cite{Suzuki15,
Yamada_JPSJ, Poenicke_PRB_1999, Zare_PRB_2008, Suzuki16}. Thus, in the
presence of disorder, the differential conductance of the $d+ip$- and
$d+is$-wave junctions may be different. Clarifying the effect of
disorder would be important for applying our theory to experimental
results.

\section{Conclusion}
\label{sec:con}
We have theoretically studied the conductance spectroscopy of
normal-metal/$d_{xy}$-wave superconductor junctions with the spin-triplet 
$p_y$-wave subdominant order at the interface utilizing 
the quasiclassical Eilenberger formalism. We have considered
the ballistic junction, where a $\delta$-function-type insulating
barrier exists at the interface. The conductance spectra are
calculated using the self-consistent pair
potentials. 

The calculated conductance spectra show that the ZBCP originating
from the $d_{xy}$-wave pair potential is not split by the $p_y$-wave
subdominant pair potential at the interface, in contrast to the $s$-wave subdominant
component, which is known to split the ZBCP. The $p_y$-wave pair potential
has nodes at $k_y=0$, which does not affect the zero-energy states
around $k_y=0$. The contributions from these channels form the ZBCP in
the conductance spectra, even in the presence of the $p_y$-wave
subdominant pair potential. 

In addition, we have studied the effect of the $p_y$-wave subdominant pair
potential on the dispersion of the surface states of the $d_{xy}$-wave SC. The $p_y$-wave
subdominant component changes the zero-energy flat band formed with
the Andreev bound states to a
V-shape or inverted V-shape, depending on the spin subspace. The
spin-subspace dependence stems from the coexistence of the
spin-singlet and spin-triplet pair potentials near the interface.

\begin{acknowledgments}
The authors would like to thank 
S.~Matsubara, S. Kashiwaya, H.~Kontani, and Y. Maeno for the useful discussions. 
This work was supported by Grants-in-Aid from JSPS for Scientific
Research on Innovative Areas ``Topological Materials Science''
(KAKENHI Grant Numbers JP15H05851, JP15H05852, JP15H05853 and JP15K21717), 
Scientific Research (A) (KAKENHI Grant No. JP20H00131), 
Scientific Research (B) (KAKENHI Grant Numbers JP18H01176 and JP20H01857), 
Japan-RFBR Bilateral Joint Research Projects/Seminars number
19-52-50026, and the JSPS Core-to-Core program ``Oxide
Superspin'' international network. 
\end{acknowledgments}

\appendix


\section{Profile of the pair potential}

\begin{figure}[tb]
	\centering
  \includegraphics[width=0.48\textwidth]{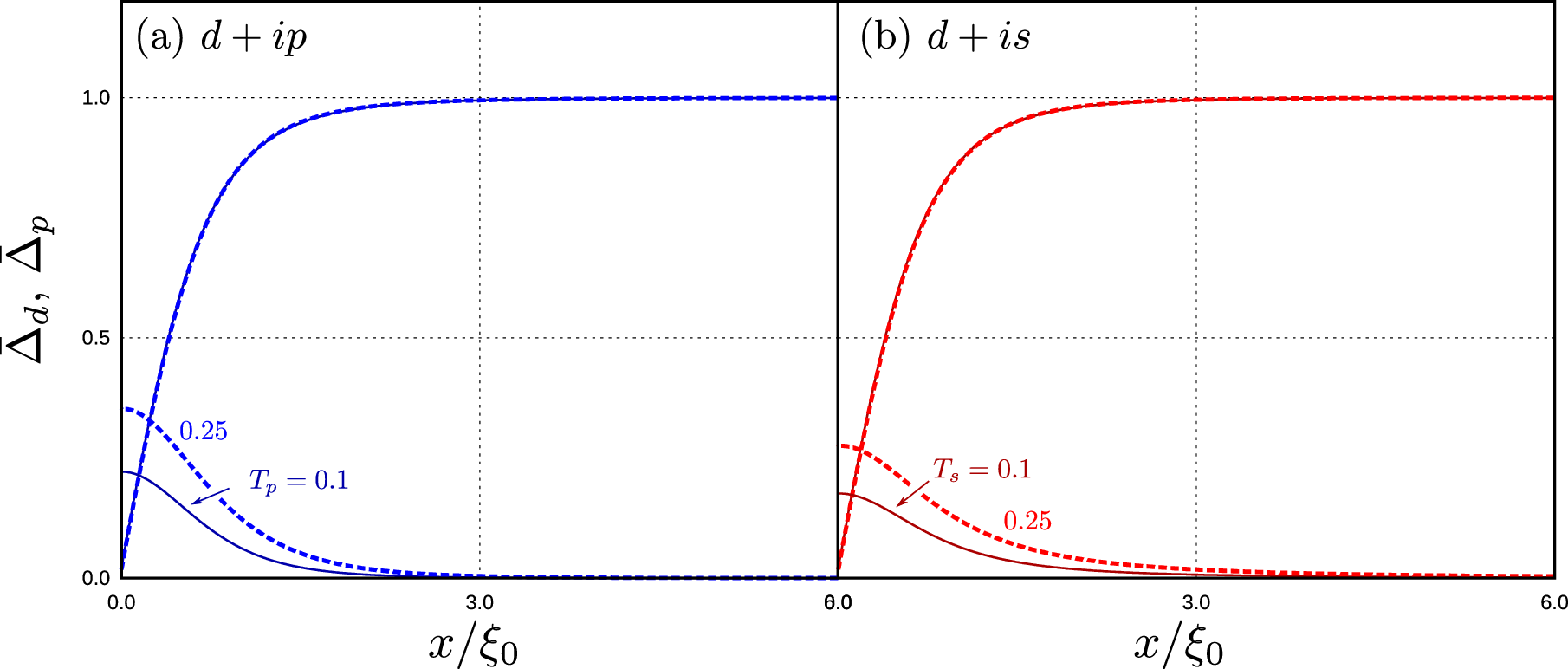}
	\caption{Spatial dependences of the pair potentials in (a) $d+ip$- and (b)
	$d+is the $-wave junctions. The parameters are set to the same values used in 
Fig.~\ref{fig:Conduc}. }
	\label{fig:pp}
\end{figure}

The profiles of the pair potentials of the $d+ip$- and $d+is$-wave
junctions are shown in Fig.~\ref{fig:pp}(a) and \ref{fig:pp}(b),
respectively. The pair potentials are normalised by that in a
homogeneous $d$-wave superconductor (i.e., $\bar{\Delta}_p = 
{\Delta}_{p(s)} / \Delta_d|_{x\to \infty}$). The
parameters are set to the same values used in 
Fig.~\ref{fig:Conduc}. The amplitude of the subdominant pair potential
depends on the effective critical temperature $T_p$ and $T_s$. The
The $p$-wave subdominant pair potential is slightly larger than the
$s$-wave one. The
subdominant component affects slightly on the $d$-wave dominant
component. However, the effect is negligible. 

\section{Spontaneous charge current}
The Eilenberger equation \eqref{eq:Eilenberger} can be rewritten as 
\begin{align}
  & i \alpha v_{F_x} \partial_x \check{g} 
	+ [\check{\mathcal{H}}, \check{g}]_- = 0, 
	\\
	& \check{\mathcal{H}}(x, \alpha, \phi, i \omega_n)
	= \left[ \begin{array}{cc}
	i \omega_n \hat{\sigma}_0 & \hat{\Delta}_\alpha  \\
	-\hat{\Delta}_\alpha^\dagger  & -i \omega_n \hat{\sigma}_0
	\end{array} \right], 
\end{align}
where $\check{g} = \check{g}_{\alpha \alpha}$. 
In this section, we make $\alpha$ explicit only when necessary. 
The current density in the $y$-direction can be obtained from the
Green's function: 
\begin{align}
  j_y(x)  
	& = e v_F N_0 \frac{\pi}{i \beta}
	\sum_{\omega_n}
	\langle 
	  k_y \tr{ \hat{g}(x, k_y, i \omega_n)}
	\rangle, 
\end{align}
where $e<0$ is the charge of a quasiparticle, $k_y = \sin \phi$,
and $\beta = 1/T$. Using
 the basic symmetry of the Green's
function $
\check{g}(x, \alpha, \phi, i \omega_n)
=
-
\check{\tau}_3
\left\{ \check{g}(x, \alpha, \phi, -i \omega_n) \right\}^\dagger
\check{\tau}_3
$, 
we can reduce the current density into the form
\begin{align}
	\frac{j_y}{j_0}
	& = \frac{T}{T_d}
	\sum_{\omega_n>0}
	\langle 
	  k_y  \mathrm{Im} \{ \tr{ \hat{g} } \}
	\rangle, 
	\\
	& = \frac{T}{T_d}
	\sum_{\omega_n>0}
	\sum_\alpha
	%
	\int_0^{\pi/2}
	  \mathrm{Im} \{ \mathrm{Tr}[
		\hat{g}(x, \alpha,  \phi, i \omega_n) 
		\notag \\
		& \hspace{26mm}
	 -\hat{g}(x, \alpha, -\phi, i \omega_n)
		] \}
    \sin \phi \, d \phi, 
	\label{eq:appen01}
\end{align}
where $j_0 = 2 \pi e v_F N_0 T_d$. In Eq.~\eqref{eq:appen01}, we have
divided the interval of integration into two regions. 

The matrices $\check{\mathcal{H}}$ for the $d_{xy}$-wave SC
can be written as 
\begin{align}
	\check{\mathcal{H}}
	& = \left[ \begin{array}{cc}
	i \omega_n \hat{\sigma}_0 & 
	 i\hat{\sigma}_2 \Delta_d   \sin (2 \phi )
	\\
	 i\hat{\sigma}_2 \Delta_d^* \sin (2 \phi )
	& 
	-i \omega_n \hat{\sigma}_0
	\end{array} \right], 
\end{align}
which satisfies the symmetry relation 
\begin{align}
\check{\mathcal{H}}(x,\alpha, \phi, i \omega_n) = 
	\check{\tau}_2 \left[ \check{\mathcal{H}}(x,\alpha, -\phi, i \omega_n) 
	\right]^* \check{\tau}_2. 
	\label{eq:appen04}
\end{align}
This relation means that the Green's
function has the symmetry in the particle-hole space
$
  \check{g}(x,\alpha, \phi, i \omega_n) = 
	\check{\tau}_2 
	\left[ \check{g}(x,\alpha, -\phi, i \omega_n) 
	\right]^* \check{\tau}_2
$, 
meaning that 
\begin{align}
  &       \hat{g}(x,\alpha,  \phi, i \omega_n) = -
	\left[ \hat{g}(x,\alpha, -\phi, i \omega_n) 
	\right]^*. 
	\label{eq:appen03}
\end{align}
Substituting Eq.~\eqref{eq:appen03} into \eqref{eq:appen01}, we can
demonstrate that no spontaneous current flows at a surface of a 
$d_{xy}$-wave SC without a subdominant pair potential. 

\begin{widetext}
The matrices $\check{\mathcal{H}}$ for the $d+ip$- and $d+is$-wave SCs
can be written as 
\begin{align}
	\check{\mathcal{H}}_p
	& = \left[ \begin{array}{cc}
	i \omega_n \hat{\sigma}_0 & 
	 i\hat{\sigma}_2 \Delta_d \sin (2 \phi )
	+\hat{\sigma}_1 i\Delta_p \sin \phi
	\\
	 i\hat{\sigma}_2  \Delta_d^* \sin (2 \phi )
	+ \hat{\sigma}_1 i\Delta_p^* \sin \phi
	& 
	-i \omega_n \hat{\sigma}_0
	\end{array} \right], 
	\\
	\check{\mathcal{H}}_s
	& = \left[ \begin{array}{cc}
	i \omega_n \hat{\sigma}_0 & 
	 i\hat{\sigma}_2 \left\{ \Delta_d \sin (2 \phi )
	+i\Delta_s \right\}
	\\
	 i\hat{\sigma}_2 \left\{ \Delta_d^* \sin (2 \phi )
	+i\Delta_s^* \right\}
	& 
	-i \omega_n \hat{\sigma}_0
	\end{array} \right]. 
\end{align}
\end{widetext}
We can show that 
the matrix $\check{\mathcal{H}}_p$ satisfies Eq.~\eqref{eq:appen04},
whereas $\check{\mathcal{H}}_s$ does not due to $\Delta_s$. 
Namely, we can demonstrate that no spontaneous charge current flows
in the $d+ip$-wave case, whereas the current flows spontaneously in
the $d+is$-wave case.

\bibliography{TKBibTK}
\end{document}